# Phase transitions in the spin-1/2 Heisenberg antiferromagnet on the square lattice

Jie Qiao[1*], Shu-Hao Zhang[2], Jing-Bo Qin[1], Xiao-Long Zhao[1], Qiang Cheng[1]

[1]School of Science, Qingdao University of Technology, Qingdao, Shandong, China
[2]School of Liberal Arts and Sciences, Qingdao Binhai University, Qingdao, Shandong, China

The nature of the intermediate ground-state phase in the spin-1/2 frustrated square lattice model has long been debated. We investigate the phase diagram of this model using cluster density matrix embedding theory. Two antiferromagnetic phases are directly identified from our analysis of the ground-state wave function: the Néel phase for   and the striped phase for  . Although our analysis of the ground state finds no direct evidence of an internal phase transition within the intermediate phase, an examination of the first excited state reveals a continuous quantum phase transition therein, with a critical point at   that separates the intermediate phase into a PVBS and a CVBS.

[*] Corresponding author's email: qiaojie@qut.edu.cn

# 1. Introduction

The exploration of quantum phases and phase transitions in strongly correlated systems has long captivated the attention of physicists [1-4]. These studies not only provide valuable insights into the fundamental behavior of matter but also reveal the exotic phenomena exhibited by these systems [5-14]. The $J_1$-$J_2$ Heisenberg model on a square lattice is a prominent example. In the early days of high-temperature superconductivity research, the $J_1$-$J_2$ model on a square lattice was one of the most important frustrated magnetic models, attracting significant research interest from both theoretical and experimental perspectives. Increasingly, experiments and theories indicate that real materials such as the vanadium layered oxides $Li_2VO(Si,Ge)O_4$ [15-17], the double perovskite compounds $Sr_2CuTe_{1-x}W_xO_6$ [18, 19], and the polycrystalline sample $BaCdVO(PO_4)_2$ [20, 21], can be adequately described by frustrated square lattice antiferromagnets with nearest-neighbor ($J_1$) and next-nearest-neighbor ($J_2$) exchanges. Frustrated square lattice systems can be described by the two-dimensional $J_1$-$J_2$ Heisenberg model, a simple yet typical quantum spin model. The antiferromagnetic $J_1$-$J_2$ model has been studied in detail by various methods due to its extreme importance in condensed matter physics.

Over the past three decades, the nature of the zero-temperature phase diagram of the spin-1/2 $J_1$-$J_2$ Heisenberg model on a square lattice has been a subject of controversy and remains a fundamental unsolved problem in quantum many-body theory. For the intermediate coupling region, it has long been believed that quantum fluctuations will destroy antiferromagnetic long-range order before the maximum frustration point of the classical model at $J_2/J_1$=0.5, potentially establishing a new paramagnetic phase. The nature of this paramagnetic phase is of great interest. Various methods have been developed to study the intermediate paramagnetic phase [22-52], and different candidate ground states have been proposed, including the columnar valence bond solid (CVBS) state [22, 24-26, 45], the plaquette valence bond solid (PVBS) state [11, 28-33, 39-41, 51, 52], and (gapped) gapless quantum spin liquid (QSL) states [11, 36, 38, 42, 43, 46-50, 53-55]. However, the physical nature of the paramagnetic phase remains a mystery. These insights have been obtained through a variety of methods, including exact diagonalization [24, 30, 56], series expansion [26, 34, 57], density matrix renormalization group (DMRG) [11, 36, 40], (infinite) projected entangled pair states ((i)PEPS) [44, 45, 54, 58, 59], neural networks [50], and quantum Monte Carlo (QMC) [47, 50, 53].

A substantial body of research has been dedicated to exploring the phase diagram of the $J_1$-$J_2$ Heisenberg model on a square lattice. This model has become a focal point in quantum magnetism research. A recent study employing the iPEPS algorithm simulated the global phase diagram of the $J_1$-$J_2$ Heisenberg model, revealing that the non-magnetic region spans $0.45 \lesssim J_2/J_1 \leq 0.61$. Within this range, it identified a gapless spin liquid phase for $0.45 \lesssim J_2/J_1 \lesssim 0.56$ and a weak valence bond solid (VBS) phase for $0.56 \lesssim J_2/J_1 \leq 0.61$ [54]. Subsequently, a very recent study using the DMRG and fully augmented matrix product states methods analyzed level crossings in the excitation spectrum to pinpoint phase transition points. This study found a direct phase transition between the Néel antiferromagnetic and VBS phases at $J_2/J_1=0.535(3)$, with no intermediate spin liquid phase present [60]. Additionally, several other works have suggested the existence of a phase transition in the intermediate region[11, 50, 53]. However, despite the abundance of results and analyses, the nature of the non-magnetic region within the range $0.4 \lesssim J_2/J_1 \lesssim 0.6$ remains a subject of intense debate and active research. The intermediate-phase phase transition is a continuous phase transition, which is challenging to detect in the ground state but observable in the first excited state [11, 49, 61, 62].

In this paper, we employ the Cluster Density Matrix Embedding Theory (CDMET) to calculate the ground state and first excited state of the spin-1/2 antiferromagnetic $J_1$-$J_2$ model. The ground-state results reveal that the transition from the Néel antiferromagnetic phase to the intermediate phase is continuous, while the transition from the intermediate phase to the stripe antiferromagnetic phase is first-order. By probing the first excited state, we identify a continuous phase transition point within the intermediate phase of the ground state at $J_2=0.55$. By calculating the correlation functions, we find a valence-bond-solid (VBS) phase transition in the first excited state of the impurity clusters. As shown in Figure 1, we present a phase diagram for a $J_1$-$J_2$ square lattice model.

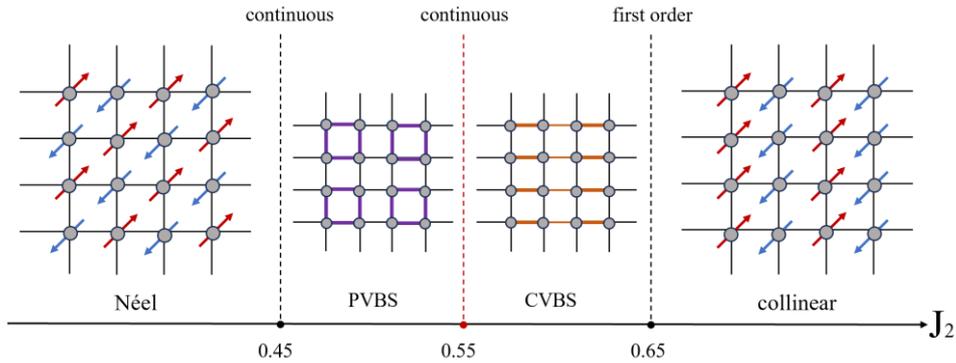

**Figure 1.** The phase diagram of spin-1/2 $J_1$-$J_2$ square-lattice Heisenberg antiferromagnetic model. The system exhibits a Néel antiferromagnetic order when $J_2<0.45$ (continuous phase transition) and transitions to a collinear antiferromagnetic phase when $J_2>0.65$ (first-order phase transition). A continuous phase transition exists at $J_2=0.55$ within the intermediate region ($0.45 \leq J_2 \leq 0.65$), splitting this phase into two distinct VBS phases.

## 2. The $J_1$-$J_2$ Model and CDMET Approach

The spin-1/2 $J_1$-$J_2$ antiferromagnetic Heisenberg model is a typical model for studying the interplay between frustration effects and quantum phase transitions. It is generally established that the ground state of this model hosts two distinct long-range ordered phases, which are separated by an intervening disordered quantum paramagnetic phase. The Hamiltonian of this model is given by the following expression

$$H = J_1 \sum_{\langle i,j \rangle} S_i S_j + J_2 \sum_{\langle\langle i,k \rangle\rangle} S_i S_k, \quad (1)$$

where $S_i$ is the spin-1/2 operator located at site i, and the summation is performed over nearest-neighbor pairs ($\langle i,j \rangle$) and next-nearest-neighbor pairs ($\langle\langle i,k \rangle\rangle$), as shown in Figure 2. This model includes nearest-neighbor ($J_1$, the energy unit in this work) and next-nearest-neighbor ($J_2$) exchange interactions, demonstrating the subtle interplay between competing magnetic interactions.

CDMET has been proven a reliable method for treating strongly correlated systems [12, 63-70]. As shown in Figure 2, we divide a finite square lattice into numerous 2×2 spin clusters and arbitrarily select one cluster as the impurity cluster, with the remaining spin clusters treated as the bath state. This size and shape are chosen to maintain the equivalence of all sites within the cluster. The 2×2 spin clusters match the $C_4$ rotational symmetry of the Néel state and the $C_2$ rotational symmetry of the collinear state. Isaev et al. show that 2×2 spin clusters are suitable for the square $J_1$–$J_2$ lattice [33]. It is clear that considering a larger spin cluster could improve the results. However, the number of bath states exponentially increases with the size of the spin cluster. Our group's previous work indicate that, unlike other simulation methods which exhibit significant size effects, the bond energy of the impurity cluster is insensitive to the lattice size [12]. This difference implies that using CDMET, we can obtain reasonable results in the thermodynamics limit by calculating only a finite lattice system.

The wave function of the 2×2 spin cluster impurity model is expressed exactly by

$$|\phi\rangle = \sum_{l=1}^{16} a_l |\mu\rangle_l |v_{bps}\rangle_l, \quad (2)$$

where the $\{|v_{bps}\rangle\}$ is 16 block-product states, which is referred to as the bath state. The approximation of the bath state is consistent with the hierarchical mean-field theory with 2×2 spin cluster. The $\{|\mu\rangle\}$ is 16 standard basis vectors and the $\{a_l\}$ represents the expansion coefficients of the wave function. We use $\|(H-E)|\phi\rangle\|^2$ to minimize the length of the state. If the wave function is accurate, the norm is zero. $H$, $\varphi$, and $E$ represent the system's Hamiltonian, the CDMET wave function, and the target energy, respectively.

Through the variational method, we can obtain an approximate eigenstate $|\phi\rangle$ which eigenvalue is closest to *E*. We use the following variational prescription:

$$\frac{\partial}{\partial \phi_{i,\sigma}} \|(H-E)|\phi\rangle\|^2 \simeq 0, \qquad (3)$$

where $\partial/\partial \phi_{i,\sigma}$ indicates the variation of the $|\phi\rangle$ by varying the real matrix $\phi_{i,\sigma}$ on site $i$ and spin $\sigma$. In our previous work, we employed this method to calculate the excited states of the $J_1$-$J_2$ Heisenberg model at different energy levels and discovered the presence of an energy gap in the intermediate phase [68]. Therefore, based on this foundation, we can gradually approximate the first excited state using the least squares method in this study. We employ an 8×8 system with periodic boundary conditions, which has been demonstrated to closely approximate an infinite system. The computation cost is lower than that of other methods. We obtained a relative error between approximately $10^{-6}$ and $10^{-5}$ from our calculations on an 8×8 lattice system [67].

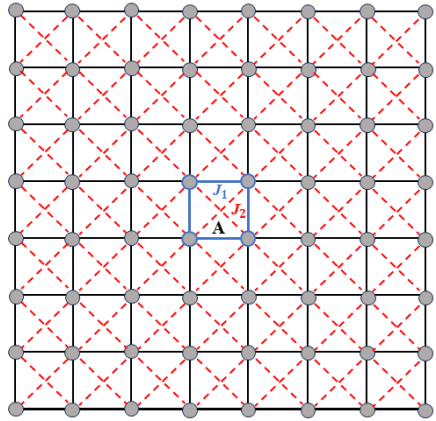

**Figure 2.** A lattice under periodic boundary conditions is divided into multiple 2×2 spin clusters, where one cluster (A) is regarded as an impurity cluster, and the remaining spin clusters are considered as the bath state.

## 3. Results and Discussion

Given the effectiveness of the von Neumann entropy in detecting phase transitions, we employ it in this study to investigate the phase transitions of the system. Discontinuities or singularities in entanglement entropy signal the onset of first-order quantum phase transitions, whereas divergent peaks in entropy derivatives characterize continuous quantum phase transitions [71-76]. We calculate the single-site and four-site von Neumann entanglement entropy between the impurity cluster and its environment, as illustrated in Figure 2. When partitioning the lattice spin system into subsystem A and its complement B, the von Neumann entanglement entropy for both subsystems can be readily obtained and is defined as

$$S(\rho_A) = -Tr(\rho_A \ln \rho_A), \qquad (4)$$

where $\rho_A = tr_B |\phi\rangle\langle\phi|$ is the reduced density matrix. The evolution of entanglement entropy for the ground state (see Figure 3) and first excited state (see Figure 5) is plotted against $J_2$. The data series are labeled as follows: $S_{GS}$ and $S_{ES}$ correspond to the single-site entanglement entropy of the ground state and the first excited state, represented by square symbols, while $S_{GP}$ and $S_{EP}$ correspond to the four-site entanglement entropy, represented by circular symbols.

*3.1. Directly Observable Phase Transitions in the Ground State*

In this subsection, we discuss the phase transitions that are directly evident in the ground state. These are the two antiferromagnetic phases, which are well known. In Figure 3, we calculate the ground-state energy of a single spin within the impurity cluster ($E_i$), entanglement entropy ($S_{GS}$ and $S_{GP}$) and its first-order derivatives ($S'_{GS}$ and $S'_{GP}$) as functions of $J_2$. As shown in Figure 3(a), the energy $E_i$ of the embedded single spin exhibits a discontinuity at $J_2=0.65$. Simultaneously, both entanglement entropy curves in Figure 3(b) display discontinuities at the same critical value $J_2=0.65$, indicating a first-order phase transition at this point. Furthermore, the first derivative of the entanglement entropy, shown in Figure 3(c), captures a second-order (continuous) phase transition, with all curves exhibiting peaks at $J_2=0.45$, confirming this as a second-order phase transition point. These results are in good agreement with the results obtained by other approximations [32, 39, 54, 77, 78]. Darradi et al. used the coupled cluster method for high orders of approximation and complementary exact diagonalization in its study [32], with results indicating that the quantum critical points for both the Néel and collinear orders are $J_2^{c1} \approx (0.44 \pm 0.01) J_1$ and $J_2^{c2} \approx (0.59 \pm 0.01) J_1$, respectively. Yu et al. applied the plaquette renormalization scheme of tensor network states and observed a continuous transition between the Néel order and the plaquettelike order near $J_2^{c1} \approx 0.40 J_1$, the collinear order emerges at $J_2^{c1} \approx 0.62 J_1$ through a first-order phase transition [39].

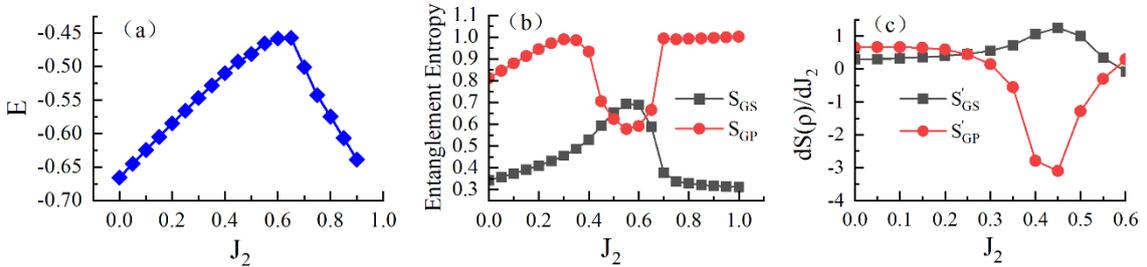

**Figure 3.** Panels (a), (b), and (c) show, respectively, the ground-state energy of a single spin within the impurity cluster ($E_i$), entanglement entropy ($S_{GS}$ and $S_{GP}$), and its first-order derivative ($S'_{GS}$ and $S'_{GP}$) as functions of $J_2$.

Furthermore, by calculating the spin expectation value components $\eta=(S_x, S_z)$, we obtain the spin vector diagrams shown in Figure 4(a) for $J_2=0.20$ and Figure 4(b) for $J_2=0.80$. When $J_2<0.45$, the spins on neighbor lattice sites are antiparallel, while those on diagonal sites are parallel, forming a staggered periodic arrangement characteristic of the Néel state, as illustrated in Figure 4(a). When $J_2>0.65$, the spins are aligned parallel along one crystallographic axis and antiparallel along the other, resulting in a collinear order periodic pattern, indicative of the stripe state, as shown in Figure 4(b).

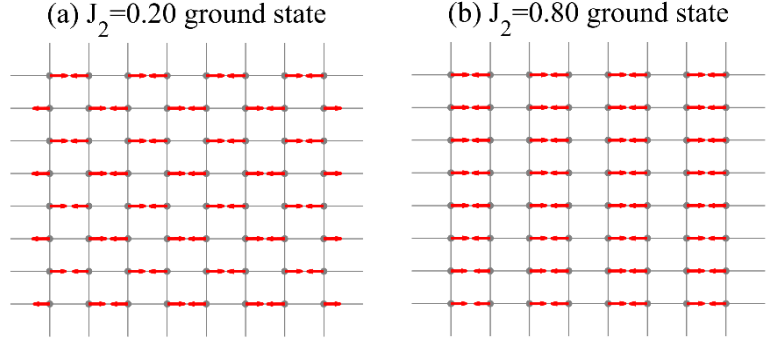

**Figure 4.** Vector map of the spin expectation value: for the ground state at (a) $J_2 = 0.20$, corresponding to the Néel state, and (b) $J_2 = 0.80$, corresponding to the striped state.

*3.2. The Hidden Phase Transitions and the First Excited State*

In recent years, numerous results have indicated that the intermediate phase exhibits a phase transition point [11, 49, 50, 53, 54, 60]. Wang et al. use the DMRG method to identify excited-level crossings versus the coupling ratio $g = J_2/J_1$, identifying a phase transition at $g_{c2} \approx 0.52$ [11]. Similarly, Qian et al. utilize excited-level crossing analysis to pinpoint the phase transition points $J_2/J_1=0.535(3)$ [60]. Liu et al. use the finite projected entangled pair state (PEPS) algorithm to determine the three phase transition points of the ground state phase as $J_2=0.45$, $J_2=0.56$, and $J_2=0.61$ [54]. Although this phase transition is not directly observed in the ground state, the first excited state can serve as a probe for it.

We calculate the first excited state in order to determine the position of the intermediate phase transition point. We present the variation curves of entanglement entropy (Figure 5(a)), and its second derivative (Figure 5(b)) with respect to $J_2$. In Figure 5(a), both the single-site entanglement entropy $S_{ES}$ and the four-site entanglement entropy $S_{EP}$ exhibit discontinuities near $J_2=0.55$. The second-order derivatives of these two entanglement entropies ($S''_{ES}$ and $S''_{EP}$) exhibit pronounced peaks at $J_2=0.55$ (see Figure 5(b)). We determine that a continuous phase transition occurs at this point. This finding is further corroborated by the recent work of Nomura et al., who explicitly demonstrate the existence of a continuous phase transition at $J_2=0.54$ [49].

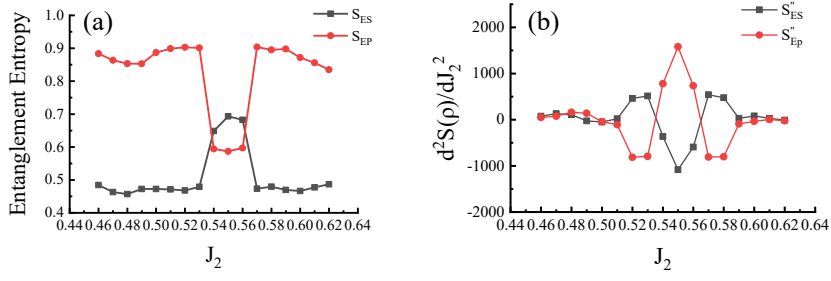

**Figure 5.** Panel (a) shows the entanglement entropy ($S_{ES}$ and $S_{EP}$) of the first excited state as a function of $J_2$. Panel (b) shows its second-order derivative ($S''_{ES}$ and $S''_{EP}$) as a function of $J_2$.

Numerous research findings have demonstrated evidence of the existence of a VBS phase in the intermediate phase of this system, lattice translation symmetry is spontaneously broken [26, 30, 33, 39-41, 45, 60]. By computing the nearest-neighbor spin correlation functions in the first excited state, we attempt to analyze the nature of the intermediate phase. We employ exactly representable impurity clusters to demonstrate this result. Figure 6 displays the corresponding correlation function values within impurity clusters (A) and between neighbor clusters for the first excited state. The spin-spin correlation function for neighbor spins is evaluated using

$$\xi = \langle S_i S_j \rangle, \qquad (5)$$

where $i$ and $j$ denote neighboring spin pairs. In Figure 6, dashed lines represent intra-cluster correlations, while solid lines denote inter-cluster correlations. Line thickness reflects the correlation strength.

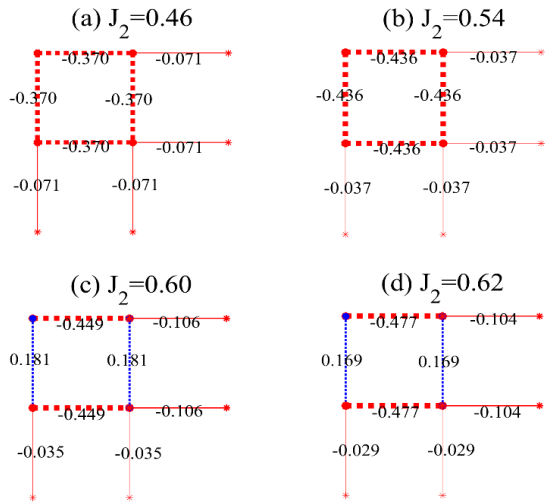

**Figure 6.** The correlation functions of the first excited state of impurity clusters (A) for different values of $J_2$. (a) $J_2 = 0.46$, (b) $J_2 = 0.54$, (c) $J_2 = 0.60$, (d) $J_2 = 0.62$. Red and blue lines represent $\xi < 0$ and $\xi \geq 0$, respectively.

As shown in Figure 6, we present the correlation functions for different values of $J_2$. When $J_2$=0.46 (Figure 6(a)) and $J_2$=0.54 (Figure 6(b)), the correlation functions between nearest-neighbor sites within the clusters are equal and both negative, with values of $\xi$=-0.370 and $\xi$=-0.436, respectively. In contrast, the correlation functions between neighboring sites across clusters are also equal but relatively small in magnitude, with values of $\xi$=-0.071 and $\xi$=-0.037, respectively. These results indicate that in the range of 0.45≤$J_2$≤0.55, each cluster acts as a strongly correlated unit, and the system exhibits a plaquette arrangement—a hallmark of the PVBS phase.

When $J_2$=0.60 (Figure 6(c)) and $J_2$=0.62 (Figure 6(d)), the correlation functions between nearest-neighbor sites within the clusters become unequal. The correlation functions along the Y-direction within the clusters turn positive (with values of $\xi$=0.181 and $\xi$=0.169, respectively), while those along the X-direction remain negative (with values of $\xi$=-0.449 and $\xi$=-0.477, respectively). Significant changes also occur in the correlation functions between neighboring sites across clusters: although still negative, the values along the X-direction become relatively large in magnitude (with values of $\xi$=-0.106 and $\xi$=-0.104, respectively), while those along the Y-direction become relatively small (with values of $\xi$=-0.035 and $\xi$=-0.029, respectively). In this work, the X-direction consistently refers to the horizontal direction of the lattice, and the Y-direction to the vertical direction. These results indicate that when 0.55<$J_2$≤0.65, the correlation functions—both within and across clusters—exhibit distinct differences between the X and Y directions. This behavior is consistent with the CVBS phase, where valence bonds form a columnar periodic arrangement along specific crystallographic directions (either vertical or horizontal).

Although our results identify the intermediate phase as a VBS, other studies suggest it could be a $Z_2$ quantum spin liquid [11, 36, 38, 42, 43, 46-50, 53-55] characterized by weak dimerization (i.e., VBS order). This implies that the true intermediate phase may either be a coexistence state of spin liquid and VBS order, or that these two states are energetically proximate.

According to Landau's phase transition theory, the occurrence of a phase transition in a system corresponds to a change in its symmetry, and a continuous phase transition is a transformation from a high-symmetry phase to a low-symmetry phase. The order parameters $F_X$ and $F_Y$ reflect the degree of bond-energy dimerization and can illustrate the change in the lattice's translational symmetry. They definition are provided in References [79, 80] as:

$$F_X = \frac{1}{N}\sum_{x,y}(-1)^x S_{x,y}S_{x+1,y}, \quad F_Y = \frac{1}{N}\sum_{x,y}(-1)^y S_{x,y}S_{x,y+1}, \qquad (6)$$

where $x$ and $y$ are coordinates of a clusters in the lattice. As shown in Figure 7, $F_X$ (denoted by a square) and $F_Y$ (denoted by a circle) represent the degree of dimerization along the lattice's X- and Y-directions, respectively.

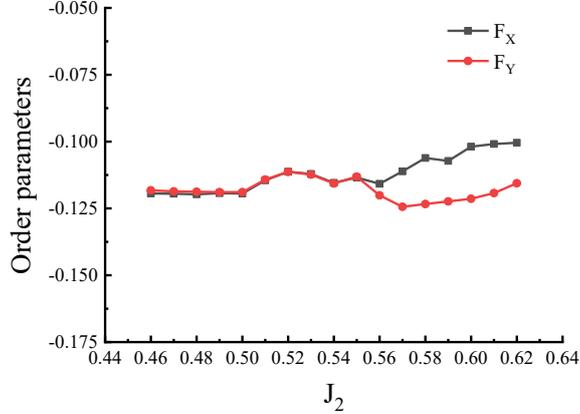

**Figure 7.** The variation of the order parameters $F_X$ and $F_Y$ with $J_2$ in the first excited state.

As shown in Figure 7, the values of order parameters $F_X$ and $F_Y$ are nearly equal over the parameter range 0.45<$J_2$≤0.55, where their absolute difference |$F_X$-$F_Y$| is negligible (less than 0.001). This indicates that the lattice retains its translational symmetry, marking the formation of an isotropic plaquette structure in the system. However, over the range of 0.55<$J_2$<0.65, the order parameters $F_X$ and $F_Y$ are markedly distinct, as their absolute difference |$F_X$-$F_Y$| is almost greater than 0.01. This indicates that the system exhibits significant anisotropy between the X and Y directions, implying the breaking of translational symmetry and the formation of a columnar arrangement. The findings regarding the symmetry change support the conclusion of a continuous phase transition at $J_2$=0.55.

## 4. Conclusion

In this study, we employ the CDMET method to compute the ground state and first excited state of the spin-1/2 $J_1$–$J_2$ Heisenberg model. Based on the analysis of the ground-state results, we identify the phase transition points for two antiferromagnetic phases, namely, the Néel phase and the striped phase. The transitions are identified as continuous at $J_2 = 0.45$ and first-order at $J_2 = 0.65$. Within the intermediate region (0.45≤$J_2$≤0.65), by examining the entanglement entropy of the first excited state as a function of $J_2$, a continuous phase transition is identified at $J_2$=0.55. Analysis of the spin-spin correlation functions and order parameters indicates the emergence of two distinct valence-bond solid (VBS) orders in this intermediate region: the plaquette VBS (PVBS) phase and the columnar VBS (CVBS) phase.

**Acknowledgements**

This work was supported by the National Natural Science Foundation of China (Grant Nos. 12005110; 12474046), and the Natural Science Foundation of Shandong Province (Grant Nos. ZR2022QA110; ZR2023MA005). The numerical calculations in this work have been done on the platform in the Supercomputing Center of Wuhan University. We acknowledge useful discussions with Quanlin Jie.

**Data Availability Statement**

The data that support the findings of this study are available from the corresponding author upon reasonable request.